# Exponential Decay of Contact-patch Friction Steering Moment with Rolling Speed

Jai Prakash*
jai.prakash@polimi.it
Mechanical Engineering Department
Politecnico Di Milano
Milan, Italy

Michele Vignati
michele.vignati@polimi.it
Mechanical Engineering Department
Politecnico Di Milano
Milan, Italy

Edoardo Sabbioni
edoardo.sabbioni@polimi.it
Mechanical Engineering Department
Politecnico Di Milano
Milan, Italy

*Presenter/Corresponding author.

*Jai Prakash,*[1] *Michele Vignati,*[2] *and Edoardo Sabbioni* [3]

# Exponential Decay of Contact-patch Friction Steering Moment with Rolling Speed

**ABSTRACT:** Steering torque is a very important quantity for the driver feeling. In fact, it gives the driver an idea of the road adherence condition during driving. Several models have been developed to simulate shear forces at the contact patch, most of them are based on semi-empirical tire models that account for slip and slip angles. They have good reliability when speed is high enough. At very low speed, like in parking, these models suffer from both reliability and numerical issues. This paper presents a model to compute the steering moment due to contact-patch friction at any longitudinal speed including pivot steering condition. In particular, it supplements the pivot steering model with a novel exponential decay of moment model to simulate for various rolling speeds of the wheel. The decay rate found to be dependent upon contact-patch geometry and rolling speed.

## Introduction

The steering system is a key part in a vehicle and its response is crucial for the driving safety of vehicles. Steering the wheels at low speed is also referred as pivot steering. The pivot steering resistance moment ( Fig. 1 ), is related to various factors such as the front axle load, the steering axis alignment parameters and steering angle of the wheel, the tire/road friction, etc. [1]–[4].

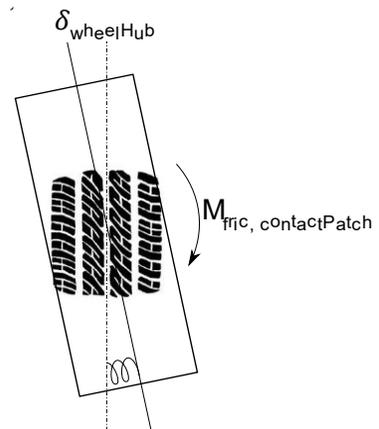

*Fig. 1: Torsion between contact patch and wheel hub*

Steering torque needed to rotate the wheels at very low speeds is an important feedback to the human driver [5]–[8]. It informs the driver about the extent of friction available at the contact of road and tires. It is also an important case in designing a vehicle steering system. It influences the choice of steering gear ratio, whether any power assistance is necessary and the nature of such power assistance. With increasing use of electrical assistance and new considerations of 'steer by wire' systems [9], [10], this subject becomes more important.

[1] Corresponding author. Doctoral candidate in Department of Mechanical Engineering,Politecnico di Milano, Italy. Email: jai.prakash@polimi.it

[2] Assistant professor in Department of Mechanical Engineering, Politecnico di Milano, Italy. Email: michele.vignati@polimi.it

[3] Associate professor in Department of Mechanical Engineering, Politecnico di Milano, Italy. Email: edoardo.sabbioni@polimi.it

At low speeds, tire lateral characteristics exhibit some peculiar phenomena, which are often overlooked but are very important in vehicle control applications, such as automatic steering systems. There is sufficient literature available which models longitudinal and lateral tire-road contact behaviour at moderate and high speeds [11]–[14]. These models face numerical problems at low speeds because longitudinal speeds of the wheel appear in the denominator of the expressions for both longitudinal and lateral slip. Empirical formulas are usually used to calculate the pivot steering resistance torque at present [15], [16], however, empirical formulas do not reflect the relationship between the steering resistance torque and the steered wheel angle.

Wang et al., [17] analyzed the effects of the steered wheel angle, friction coefficient, tire pressure and vertical load on the pivot steering resistance torque based on a vehicle road test. In their work [18], they also considered the contact between the tire and pavement in their research on the pivot steering resistance torque. However, the effect of the kingpin offset on the steering resistance torque is neglected in the modeling. Also, the analysis is performed at only pivot steering condition not during any rolling of wheels. Zhuang [19], introduced the LuGre friction model in their research on the pivot steering resistance torque, and the parameter identification and simulation calculation was carried out.

Cao et al., [20] developed a finite element model of tire-road to simulate low-speed steering friction force and to obtain a relation between steering friction coefficient with the longitudinal speed of the tire. However, this relation lacks in describing the decaying of some portion of steering moment as tire rolls.

Jagt [21], introduces a model that is especially aimed at the generation of a steering moment response at zero speed. Qualitatively good results have been obtained using this model, under a sinusoidal steer input and periodic attainment of full sliding at tire-road contact. However, it lacks in modelling variation of steering moment with rolling speed of the tire.

This paper proposes an exponential decay model to replicate the decaying behaviour of steering moment as the steered wheel rolls. Pivot steering causes twist in tire about the vertical axis. Twist is stored in form of potential energy due to tire torsion. The twist builds up until the available friction limit at the contact patch. After reaching friction limit the contact-patch starts sliding. This causes saturation in the twist, where twist is the difference in hub steering angle and contact patch steering angle. As tire starts rolling, the energy stored in tire torsion decays. The faster the tire rolls, faster the twist decays. This paper discusses a novel relation between rolling speed and decay in steering moment. The outline of the paper is as follows. The 'Method' section first derives the expression of maximum steering moment available due to friction at tire-road contact, and then uses brush model approach to derive exponential decay model for the steering moment. The 'Results' section presents the results obtained with the new model. The 'Discussion and conclusion' section discusses the key points observed in the proposed model and concludes the paper.

**Method**

I.   Friction limit

The key factors responsible for steering moment generation at very low speed mainly consists of contact patch friction, aligning moment due to kingpin inclination, caster, and camber. Contribution of contact patch friction is dominant, so effect of aligning moment due other small inclinations is neglected.

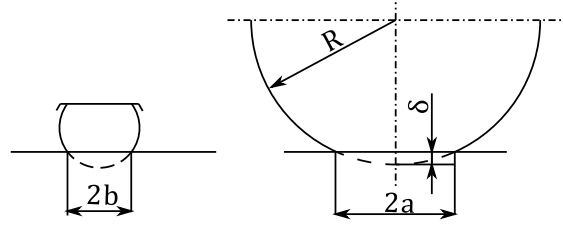

Fig. 2: Contact between tire and road surface

The surface contact between tire and road surface can be approximated as shown in Fig. 2.

Here, $\delta$ is the deformation or tire under the influence of normal load, $F_z$. $R$ is the undeformed radius of tire. $2a$ and $2b$ are length and width of tire-road contact patch, respectively. Empirical relations given in [22], can be used to calculate values of $\delta$ (cm), $2a$ (m) and $2b$ (m).

$$\delta = ck\frac{F_z^{0.85}}{B^{0.7}D^{0.43}P^{0.6}} \qquad (1)$$

$$a = \frac{D}{100}\left(\frac{\delta}{D}\right)^s \qquad (2)$$

$$b = \frac{B}{200}\left(1 - e^{-t\delta}\right) \qquad (3)$$

$c$, a parameter related to tire type. $c = 1.15$ for bias tire and $c = 1.5$ for radial tire.
$B$, width of tire in $cm$.
D, Diameter of tire in $cm$.
$k = 0.015B + 0.42$.
$P$, tire pressure with unit of $100kPa$.
$F_z$, vertical load with unit of $10N$.
$s = 0.557$, an empirical coefficient.
$t = 122.7$, an empirical coefficient.

Usually, wheel does not turn about the center of the contact patch, but it turns about its steering axis. The steering axis is different from the vertical axis passing through the contact-patch center due to scrub radius, caster, camber, and kingpin inclination angle. But, for this derivation purpose steering axis is considered as the vertical axis passing through the center of the contact-patch. When load is small the contact patch is nearly circular, but as load increases contact patch can be approximated with a rectangular shape. Fig. 3 shows considered contact patch geometry and normal pressure distribution along longitudinal axis.

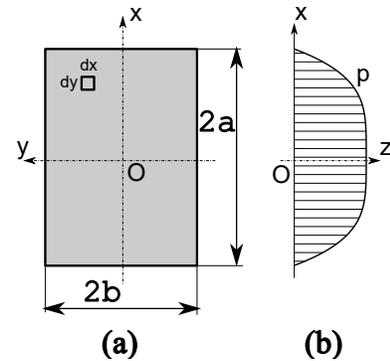

Fig. 3: Contact patch view: (a) Contact patch approximation; (b) Normal pressure distribution along longitudinal axis

Cao et al. [20], suggest $4^{th}$ order pressure distribution along longitudinal axis for radial tires.

$$p = \frac{5F_z}{8a}\left[1 - \left(\frac{x}{a}\right)^4\right]$$
(4)

Constant pressure distribution is considered along the y-axis. Infinitesimal element $dxdy$ in Fig. 3a experiences a normal load given by:

$$dF_z = \frac{p(x)}{2b} \cdot dx \cdot dy$$
(5)

As wheel steers at standstill, torsional sliding occurs at the contact patch. If dynamic friction coefficient is considered a constant μ, then friction force at the infinitesimal element would be:

$$dF_f = \mu \cdot dF_z$$
(6)

Friction moment generated around the rotation axis is:

$$dM_f = \sqrt{x^2 + y^2} \cdot dF_f$$
(7)

If the static and dynamic friction coefficient are considered equal, then integration of $dM_f$ over the contact-patch region yields the steering friction torque limit of a single tire:

$$M_{f,lim} = \int_{-a}^{a} \int_{-b}^{b} dM_f$$
(8)

II. Exponential decay

To develop the decaying expression for steering moment, a simple case of tire torsion is considered. In which at zero rolling speed, the tire is first twisted under the influence of a steering moment. Then the tire is allowed to roll in forward direction along the wheel hub plane. For the sake of simplicity, free rolling is considered along the wheel hub plane.

The brush model approach is used to derive the expression for decay in steering moment. In brush model, a row of elastic bristles touches the road plane. The bristles can deflect in any direction parallel to road surface. Considering small dimension of bristles compare to the contact patch size, it is reasonable to assume that even for small steer most the bristles would attain a state of shear stress saturation. To present the approach, Fig. 4a shows 12 bristles saturated as per the normal pressure on them. As the tire rolls a distance of $s$, bristles move in backward direction, Fig. 4b. The top half bristles carry their previous deflections, and hence their previous shear stresses. When bristle '1' traverses the distance $s$ backwards, it becomes capable of withstanding more shear stress due to increase in normal pressure. But, as the wheel is just rolling and not steering anymore, it would retain its original stress. On the contrary, the bottom half bristles lose some of its shear stresses as normal pressure distribution decreases in backward direction, as shown in Fig. 4b. As a result, in bottom half of the contact-patch, shear stress starts after $x = -s$. Fig. 4c shows remaining shear stress distribution after wheel rolled a distance $s$.

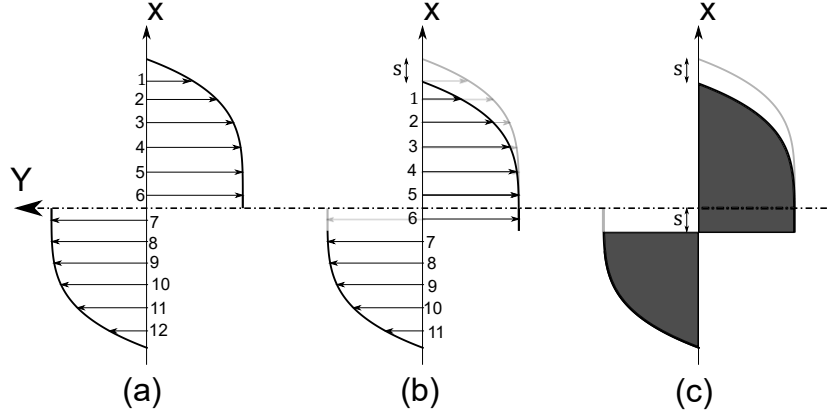

*Fig. 4: Shear stress at contact patch: (a) Saturated twisting of bristles; (b) Tire moved longitudinally considering free rolling; and (c) Remaining shear stress after the move*

In case of Fig. 4a, when the shear stress distribution is same in top and bottom half of the contact-patch. Moment per unit width is given by:

$$M_{contactpatch} = M_{top} + M_{bottom} = 2\int_0^a \mu p \cdot x \cdot dx \qquad (\,9\,)$$

where p is given by equation ( 4 ). It can be observed that in the bottom region of the contact-patch, shear stress distribution is mirrored around $x$-axis, Fig. 4a. In equation ( 9 ), $M_{contactpatch}$ is just twice of $M_{top}$. With the substitution of $p$ and then integrating equation ( 9 ), $M_{contactpatch}$ can be defined as follow.

$$M_{contactpatch} = \frac{5a\mu F_z}{12} \qquad (\,10\,)$$

As Fig. 4c summarizes, top half shear stress distribution is shifted backwards by distance $s$. But the bottom distribution remains the same. Just the integration limits are changed in bottom half shear stress distribution.

$$M_{top} = \mu \int_{-s}^{a-s} \frac{5F_z}{8a}\left[1 - \left(\frac{x+s}{a}\right)^4\right] \cdot x \cdot dx \qquad (\,11\,)$$

$$= \frac{5\mu F_z}{8a^5}\left[\frac{a^4(a-s)^2}{2} - \frac{(a-s)^6}{6} - \frac{4(a-s)^5 s}{5} + \frac{s^6}{30} - \frac{3(a-s)^4 s^2}{2} - \frac{4(a-s)^3 s^3}{3} - \frac{(a-s)^2 s^4}{2} - \frac{a^4 s^2}{2}\right] \qquad (\,12\,)$$

$$M_{bottom} = \mu \int_{-a}^{-s} \frac{5F_z}{8a} \left[ \left( \frac{x}{a} \right)^4 - 1 \right] \cdot x \cdot dx \qquad (13)$$

$$= \frac{5\mu F_z}{8a^5} \left[ \frac{a^6}{3} - \frac{a^4 s^2}{2} + \frac{s^6}{6} \right] \qquad (14)$$

$$M_{contactpatch} = M_{top} + M_{bottom} = f(s) \qquad (15)$$

Equation ( 15 ), gives the steering moment remaining after the tire rolled a distance $s$. A quick sanity check of putting $s = 0$ in equation ( 15 ) yields to equation ( 10 ), assures the correctness of the equation near the initial condition. It is interesting to know the rate of change of $M_{contactpatch}$ with rolled distance, $s$. Given that equation ( 15 ) is non-linear with $s$, the decay rate is only calculated near the initial condition, i.e., when $s$ is infinitesimal.

$$\left. \frac{dM_{contactpatch}}{ds} \right|_{s \to 0} = -\frac{\mu F_z}{2} \qquad (16)$$

Dividing equation ( 16 ) by equation ( 10 ), yields to

$$\left. \frac{dM_{contactpatch}}{ds} \right|_{s \to 0} = -\frac{1}{5a/6} M_{contactpatch} \qquad (17)$$

As $ds = V dt$, where $V$ is tire rolling velocity, gives below relation

$$\left. \frac{dM_{contactpatch}}{dt} \right|_{s \to 0} = -\frac{V}{5a/6} M_{contactpatch} \qquad (18)$$

Equation ( 18 ) indicates exponential decay of steering moment as tire rolls. Decay rate is inversely proportional to contact patch length, if contact patch is smaller than the decay is faster. Decay rate is directly proportional to rolling speed, if the rolling speed increases then the steering moment decays faster.

As shown in Fig. 5, if wheel hub is considered rigid and $K_{torsion}$ is tire torsional stiffness, then the torsion moment is given by:

$$M_{contactpatch} = -K_{torsion}(\delta_{wheelHub} - \delta_{contactPatch}) \qquad (19)$$

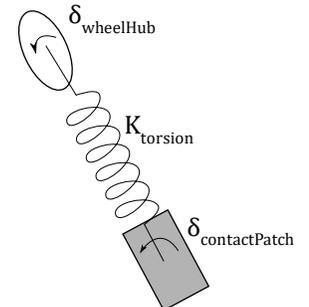

Fig. 5: Tire torsion during wheel steer

Differentiation of equation ( 19 ) w.r.t time results:

$$\frac{dM_{contactpatch}}{dt} = -K_{torsion} \frac{d}{dt}(\delta_{wheelHub} - \delta_{contactpatch}) \qquad (20)$$

Substituting equations ( 19 ) and ( 20 ) into ( 18 ) gives:

$$\frac{d}{dt}\left(\delta_{wheelHub} - \delta_{contactpatch}\right) = -\frac{V}{5a/6}\left(\delta_{wheelHub} - \delta_{contactpatch}\right) \tag{21}$$

## III. Van Der Jagt model

To test above derived model, Jagt [8] model is used as a reference model. Jagt model deals with steering moment response at zero rolling speed. The principle of Jagt approach is that at a given rate of change of the steer input the growth rate of the tire angular deflection, β̇, decreases in proportion to a function of the remaining difference between the maximum achievable deflection and the current deflection. The steering moment gradually approaches to its maximum value, which is the friction limit available at the tire-road contact-patch.

According to Jagt model, steering moment at zero rolling speed can be calculated with following set of equations:

$$\dot{\beta} = -\left[1 - p\left(\frac{M_{contactPatch}}{M_{f,lim}}\right)^2\right]\dot{\delta}_{wheelHub} \tag{22}$$

$$\dot{M}_{contactPatch} = K_{torsion} \cdot \dot{\beta} \tag{23}$$

$$p = 0; \quad if\ sgn(\beta) \neq -sgn(\dot{\delta}_{wheelHub}), \quad else\ p = 1 \tag{24}$$

To clarify, equations ( 20 ) and ( 23 ) are identical as,

$$\beta = \delta_{contactPatch} - \delta_{wheelHub} \tag{25}$$

For a standing tire (size P205/65R15) with zero rolling speed, the response to alternating steer angle variations will follow a course similar to Fig. 6. The model performance is quite good, except for a small deviation from the experimental data at the initial rising branch.

The next section elaborates fusion of Jagt model with exponential decay of steering moment model to estimate steering moment at different rolling speed.

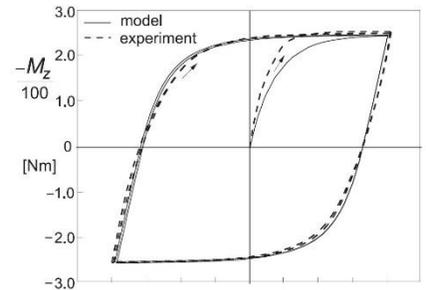

*Fig. 6: Calculated and experimentally assessed variation of the moment vs steer angle for a non-rolling tire pressed against a flat plate at a load Fz = 4800 N. Reprinted from "Tire and Vehicle Dynamics" [23]*

**Results**

Jagt model delivers good results notably when a sinusoidal steer angle variation is imposed, and the state of almost full sliding attained periodically. Pacejka [23] combines the Jagt model with tire magic formula with the help of a coefficient $w_{Vlow}$ which is given by,

$$
\begin{aligned}
if \quad & |V| < V_{low} \\
& w_{Vlow} = 0.5\left\{1 + \cos\left(\pi\frac{V}{V_{low}}\right)\right\} \\
else \quad & \\
& w_{Vlow} = 0
\end{aligned}
$$

$(26)$

where, $V_{low}$ is the rolling speed beyond which moment from the Jagt model considered vanished. The combined equation is a weighted average of moment from the Jagt model and aligning moment from the tire magic formula. Weights are $w_{Vlow}$ and $(1 - w_{Vlow})$ respectively.

In this paper, instead of opting the weighting approach to Jagt model, it is supplemented with the novel exponential decay of steering moment model. The resulting model is summation of contribution of Jagt model equation ( 22 ) and contribution of exponential decay model equation ( 21 ).

$$
\dot{\beta} = -\left[1 - p\left(\frac{M_{contactPatch}}{M_{f,lim}}\right)^2\right]\dot{\delta}_{wheelHub} - \frac{V}{5a/6}\beta
$$

$(27)$

$$
\dot{M}_{contactPatch} = K_{torsion} \cdot \dot{\beta}
$$

$$
p = 0; \quad if \; sgn(\beta) \neq -sgn(\dot{\delta}_{wheelHub}), \quad else \; p = 1
$$

$(28)$

Fig. 7, presents the counter steering moment calculated by integrating equations of motion given by equation ( 27 ) and equation ( 28 ) for various rolling speed of tire. Dynamic inputs are $\delta_{wheelHub}$, $\dot{\delta}_{wheelHub}$ and $V$. $\delta_{wheelHub}$ is a sinusoidal steering input, $\dot{\delta}_{wheelHub} = 5\sin(2\pi0.5t)$ for $V = 0, 1 \; and \; 10 \; km/h$. At high rolling speeds, steering movement is often very small. That's why, steer of an amplitude of $1 \; deg$ is given at $100 \; km/h$ speed. It can be observed that at zero rolling speed, only Jagt

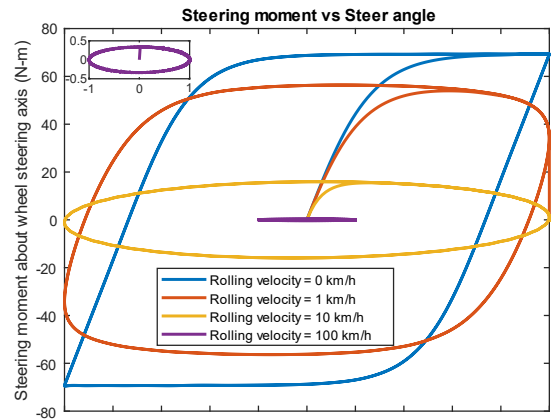

*Fig. 7: Counter steering moment for sinusoidal steering input at various speeds*

model contribution is active. As rolling speed increases, exponential decay contribution dominates, which lowers steering moment.

The parameter values used for the simulation are given in Table I.

Fig. 8**Error! Reference source not found.** presents the output of this model for a combination of pivot steering, acceleration, and then constant rolling speed of $10\ km/h$. This model is only intended to simulate torsional shear stress condition inside the contact-patch. Moment due to lateral shear stresses at the contact-patch is not accounted in this model. So, the rolling acceleration is considered as quasi-static acceleration by this model.

Fig. 8 can be divided into four sections. $0-2\ sec$ where sinusoidal steering input is given at pivot steering condition. At Pivot steering condition, steering moment saturates as per maximum friction available at the contact-patch, equation ( 8 ). The steering moment changes its sign before the steer angle does the same, mainly because the rate of change of steer angle changes its sign. Second section $2-3.5\ sec$ where rolling speed is increasing, and decay of steering moment is observed. Faster decay can be observed as rolling speed increases. Third section $3.5-4.5\ sec$ where rolling speed is constant with sinusoidal steering input. In third section, steering moment is much lower compared to moment at pivot steering because of rolling speed of $10\ km/h$. Fourth section $4.5-6\ sec$ where both rolling speed and steer input are constant. Due to exponential decay model, the steering moment vanishes ultimately. Disappearance of steering moment indicates disappearance of potential energy, which was stored inside the tire in the form of torsion.

*Table I: Parameter values for simulation*

| Parameter | Value | Unit |
|---|---|---|
| $F_z$ | 200 | $Kg$ |
| $K_{torsion}$ | 3000 | $N-m$ |
| $B$ | 14.5 | $cm$ |
| $D$ | 50.42 | $cm$ |
| $P$ | 2.5 | $100\ kpa$ |
| $\mu$ | 0.7 | [1] |
| $a\ (caluated)$ | 0.0684 | $M$ |
| $b\ (caluated)$ | 0.0725 | $M$ |
| $M_{f,lim}\ (caluated)$ | 69.28 | $N-m$ |

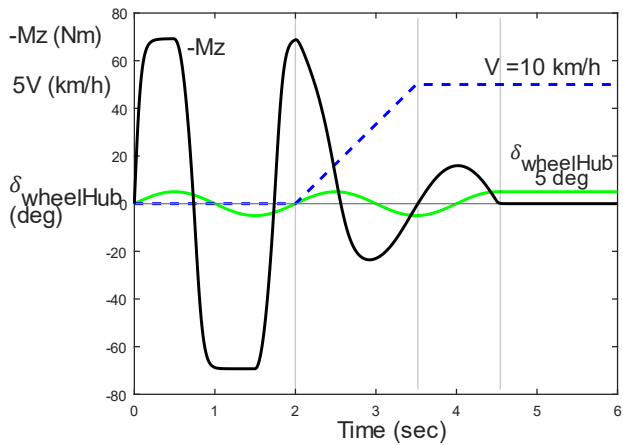

*Fig. 8: Counter steering moment for sinusoidal steering input for pivot steering, acceleration and constant non-zero rolling speed*

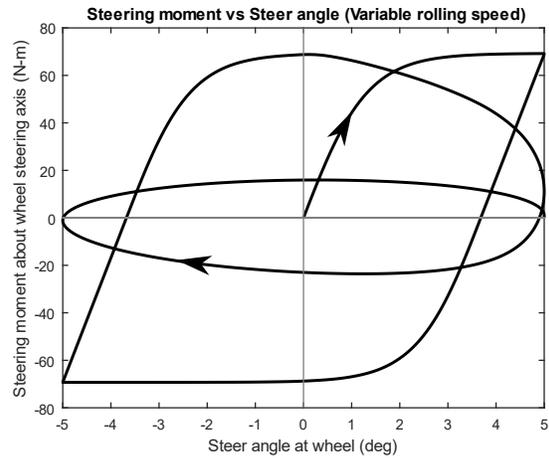

*Fig. 9: Counter steering moment vs steer angle*

The loops shown in Fig. 9, give a nice impression of the transition discussed for Fig. 8 i.e., transition from the pivot-steering to the condition of higher rolling speeds with sinusoidal steering input.

## Discussion and conclusion

In steering moment response many factors contribute. Some factors are steering axis geometry, scrub radius, tire-road contact friction, normal load, tire size, steer rate and tire rolling speed are the primary factors. If the steering axis geometry and scrub radius is neglected, then the steering moment is only generated due to lateral and torsional forces during wheel steering. Empirical model approach such as Pacejka [24] tire model can efficiently be used to model shear forces at medium to high longitudinal speeds. At low speeds, these empirical models give low values for (lateral) shear forces. Which doesn't match with the reality of high steering moment requirement for pivot steering. The proposed combination of Jagt model and exponential decay of steering moment model, outputs steering moment due to torsional stresses at contact-patch, which can directly be sum up with moment due to (lateral) shear forces given by Pacejka tire model.

Exponential decay rate of steering moment is found directly proportional to rolling speed and inversely proportional to contact patch length. This behaviour aligns with the reality, as moment observed at pivot steering decreases as wheel starts rolling. The decay rate suggested in this paper is valid only if rolled distance in each time-step is very small compared to contact-patch length, equation ( 18 ). It indicates that while performing integration of the equation of motions, time-step should be small for high rolling speed.

*Future work*: Consideration of combined slip condition to estimate friction coefficient in equation ( 6 ). As in presence of longitudinal and lateral shear stresses, friction available for torsional stresses at tire-road contact is less as compared to when they are absent. Combined slip condition will limit $\mu$, and hence $M_{lim}$ in equation ( 8 ) will be reduced. Limited $M_{lim}$ means earlier saturation of steering moment in presence of longitudinal and lateral shear stresses. Experimental validation of this novel model on tire test rig is also included in the future work.

*Applications* of this model could be many. E.g., steering feedback to human operator in case of vehicle teleoperation, estimating correct torque requirement in case of designing steer-by-wire and electric power steering systems.

# References

[1] T. Shiiba and W. Murata, "Experimental validation of steering torque feedback simulator through vehicle running test," *J. Mech. Sci. Technol.*, vol. 23, no. 4, pp. 954–959, 2009, doi: 10.1007/s12206-009-0320-9.

[2] D. Karimi and D. Mann, "Torque feedback on the steering wheel of agricultural vehicles," *Comput. Electron. Agric.*, vol. 65, no. 1, 2009, doi: 10.1016/j.compag.2008.07.011.

[3] W. Y. Park, S. Y. Kim, C. H. Lee, D. M. Choi, S. S. Lee, and K. S. Lee, "The Effect of Ground Condition, Tire Inflation Pressure and Axle Load on Steering Torque," *J. Biosyst. Eng.*, vol. 29, no. 5, 2004, doi: 10.5307/jbe.2004.29.5.419.

[4] Y. JIN, Y. LU, J. GONG, Z. LU, W. LI, and J. WU, "Design and Experiment of Electronic-hydraulic Loading Test-bed Based on Tractor's Hydraulic Steering By-wire," *Asian Agric. Res.*, vol. 7, no. 12, pp. 86–89, 2015, doi: 10.22004/AG.ECON.240745.

[5] F. Friedrichs and B. Yang, "Drowsiness monitoring by steering and lane data based features under real driving conditions," in *European Signal Processing Conference*, 2010, pp. 209–213, Accessed: Aug. 25, 2021. [Online]. Available: https://ieeexplore.ieee.org/abstract/document/7096521.

[6] D. Toffin, G. Reymond, A. Kemeny, and J. Droulez, "Role of steering wheel feedback on driver performance: Driving simulator and modeling analysis," *Veh. Syst. Dyn.*, vol. 45, no. 4, pp. 375–388, 2007, doi: 10.1080/00423110601058874.

[7] Tyler M Grant, Davood Karimi, and Danny D Mann, "Effect of Steering Torque Feedback on Driving Performance in a Tractor Simulator," Nov. 2013, doi: 10.13031/2013.24175.

[8] J. P. Switkes, J. C. Gerdes, G. F. Schmidt, and M. Kiss, "Driver response to steering torque disturbances: A user study on assisted lanekeeping," *IFAC Proc. Vol.*, vol. 40, no. 10, pp. 243–250, Jan. 2007, doi: 10.3182/20070820-3-US-2918.00034.

[9] Y. Yao, "Vehicle Steer-by-Wire System Control," *SAE Tech. Pap.*, Apr. 2006, doi: 10.4271/2006-01-1175.

[10] P. Yih and J. C. Gerdes, "Steer-by-wire for vehicle state estimation and control," in *Proceedings of AVEC*, 2004, pp. 785–790, Accessed: Aug. 25, 2021. [Online]. Available: https://citeseerx.ist.psu.edu/viewdoc/download?doi=10.1.1.520.6138&rep=rep1&type=pdf .

[11] F. Cianetti, L. Fabellini, V. Formica, and F. Ambrogi, "Development and validation of a simplified automotive steering dynamic model," *Proc. Inst. Mech. Eng. Part D J. Automob. Eng.*, vol. 235, no. 8, 2021, doi: 10.1177/0954407020984668.

[12] K. T. R. van Ende, F. Küçükay, R. Henze, F. K. Kallmeyer, and J. Hoffmann, "Vehicle and steering system dynamics in the context of on-centre handling," *Int. J. Automot. Technol.*, vol. 16, no. 5, 2015, doi: 10.1007/s12239-015-0076-4.

[13] P. E. Pfeffer, M. Harrer, and D. N. Johnston, "Interaction of vehicle and steering system regarding on-centre handling," *Veh. Syst. Dyn.*, vol. 46, no. 5, 2008, doi: 10.1080/00423110701416519.

[14]    P. E. Pfeffer and M. Harrer, "On-centre steering wheel torque characteristics during steady state cornering," 2008, doi: 10.4271/2008-01-0502.

[15]    H. Jiang, Z. Zhou, Y. Qian, and H. Liu, "Variable assist characteristics and control strategies for ECHPS in terms of maneuverability and energy efficiency," *Jixie Gongcheng Xuebao/Journal Mech. Eng.*, vol. 51, no. 22, pp. 88–97, Nov. 2015, doi: 10.3901/JME.2015.22.088.

[16]    R. S. Sharp and R. Granger, "On car steering torques at parking speeds:," *http://dx.doi.org/10.1177/095440700321700202*, vol. 217, no. 2, pp. 87–96, Aug. 2016, doi: 10.1177/095440700321700202.

[17]    Y. Wang, X. Gao, and X. Zhang, "Static steering resisting moment of tire for heavy multi-axle steering vehicle," *Nongye Gongcheng Xuebao/Transactions Chinese Soc. Agric. Eng.*, vol. 26, no. 10, 2010, doi: 10.3969/j.issn.1002-6819.2010.10.024.

[18]    W. Yunchao, P. Feng, W. Pang, and M. Zhou, "Pivot steering resistance torque based on tire torsion deformation," *J. Terramechanics*, vol. 52, no. 1, 2014, doi: 10.1016/j.jterra.2014.02.003.

[19]    Z. Ye and G. Konghui, "Tire spot turn model based on LuGre model," *Automob. Technol.*, vol. 7, 2008.

[20]    D. Cao, B. Tang, H. Jiang, C. Yin, D. Zhang, and Y. Huang, "Study on low-speed steering resistance torque of vehicles considering friction between tire and pavement," *Appl. Sci.*, vol. 9, no. 5, 2019, doi: 10.3390/app9051015.

[21]    P. Jagt van der, "The road to virtual vehicle prototyping : new CAE-models for accelerated vehicle dynamics development," Technische Universiteit Eindhoven, 2000.

[22]    J. D. Zhuang, "Calculation of Vehicle Terramechanics," in *China Machine Press: Beijing, China*, 2002, pp. 52–60.

[23]    H. B. Pacejka, "Chapter 9 - Short Wavelength Intermediate Frequency Tire Model," in *Tire and Vehicle Dynamics (Third Edition)*, Third Edition., H. B. Pacejka, Ed. Oxford: Butterworth-Heinemann, 2012, pp. 441–442.

[24]    H. B. Pacejka, "Chapter 4 - Semi-Empirical Tire Models," in *Tire and Vehicle Dynamics (Third Edition)*, Third Edition., H. B. Pacejka, Ed. Oxford: Butterworth-Heinemann, 2012, pp. 149–209.

**List of Figures Captions**

Fig. 1: Torsion between contact patch and wheel hub

Fig. 2: Contact between tire and road surface

Fig. 3: Contact patch view: (a) Contact patch approximation; (b) Normal pressure distribution along longitudinal axis

Fig. 4: Shear stress at contact patch: (a) Saturated twisting of bristles; (b) Tire moved longitudinally considering free rolling; and (c) Remaining shear stress after the move

Fig. 5:  Tire torsion during wheel steer

Fig. 6: Calculated and experimentally assessed variation of the moment vs steer angle for a non-rolling tire pressed against a flat plate at a load Fz = 4800 N. Reprinted from "Tire and Vehicle Dynamics"

Fig. 7: Counter steering moment for sinusoidal steering input at various speeds

Fig. 8: Counter steering moment for sinusoidal steering input for pivot steering, acceleration and constant non-zero rolling speed

Fig. 9: Counter steering moment vs steer angle

**List of Tables**

Table I: Parameter values for simulation